\documentclass[5p,twocolumn]{elsarticle}

\usepackage{graphicx}
\usepackage{dcolumn}
\usepackage{pifont}
\usepackage{bm}
\usepackage{multirow}
\usepackage{amsmath}
\usepackage{float}
\usepackage{txfonts}
\usepackage[version=3]{mhchem}

\usepackage[usenames,dvipsnames]{xcolor}
\definecolor{blue}{RGB}{0,0,225}
\definecolor{cream}{RGB}{222,217,201}
\definecolor{red}{RGB}{225,0,0}

\def\mathbi#1{\textbf{\em #1}}

\journal{arXiv}

\begin{document}
\title{Influence of Ti/V Cation-Exchange in Na$_2$Ti$_3$O$_7$ on Na-Ion Negative Electrode Performance: an Insight from First-Principles Study}

\author[kimuniv-m]{Chol-Jun Yu\corref{cor}}
\cortext[cor]{Corresponding author}
\ead{cj.yu@ryongnamsan.edu.kp}
\author[kimuniv-m]{Suk-Gyong Hwang}
\author[kimuniv-m]{Yong-Chol Pak}
\author[kimuniv-m]{Song-Hyok Choe}
\author[kimuniv-m]{Jin-Song Kim}
\author[kimuniv-m]{Kum-Chol Ri}

\address[kimuniv-m]{Chair of Computational Materials Design, Faculty of Materials Science, Kim Il Sung University, Ryongnam-Dong, Taesong District, Pyongyang, Democratic People's Republic of Korea}

\begin{abstract}
Sodium-titanate \ce{Na2Ti3O7} (NTO) is regarded as a highly promising anode material with a very low voltage for Na-ion batteries and capacitors, but suffered from relatively low specific capacity and poor electron conductivity.
Here we report a first-principles study of electrochemical properties of NTO and its vanadium-modified compounds, \ce{Na2Ti2VO7} and \ce{Na2TiV2O7} (NTVO), offering an insight into their detailed working mechanism and an evidence of enhancing anode performance by Ti/V cation exchange.
Our calculations reveal that the specific capacity can increase from 177 mAh g$^{-1}$ in NTO to over 280 mAh g$^{-1}$ in NTVO when using \ce{NaTi_{3-$x$}V_{$x$}O7} ($x$ = 1, 2) as a starting material for Na insertion due to higher oxidation state of \ce{V^{+5}}, together with lower voltages and small volume expansion rates below 3\%.
With Ti/V exchange, we obtain slightly higher activation energies for Na ion migrations along the two different pathways, but find an obvious improvement of electronic transport in NTVO.
\end{abstract}

\maketitle

\section{Introduction}
Ever increasing demand for a better welfare in daily life has drawn an explosive production of portable electronic devices and electric vehicles.
Together with the urgent need for exploiting renewable and clean energy sources, which are often intermittent, this requires to develop advanced electrochemical energy storage (EES) systems with high efficiency, stability and reliability.
Primarily two types of devices exist for reversible EES; secondary batteries, and electrochemical supercapacitors.
The former delivers a high energy density, while the latter provides a high power density with a longer life time.
Up till now, Li-ion batteries (LIBs) have been dominating EES markets, but recently, a great deal of renewed research attention has been paid to Na-ion based technology, such as Na-ion batteries (NIBs)~\cite{JYHwang17csr} and Na-ion capacitors (NICs)~\cite{Ding18cr}.
Due to the huge abundance and low price of sodium resources, the use of sodium instead of lithium is expected to overcome the probable shortage of lithium resources and to be suitable choice for the large-scale applications.

The key issue in realizing commercially viable NIBs and NICs is to search their proper electrode materials, especially anodic materials.
Although graphite is a common anode material in LIBs, it does not properly intercalate Na$^+$ ion due to its larger ionic radius (1.02 \AA) compared to Li$^+$ ion (0.76 \AA).
In this status, ternary graphite intercalation compounds with organic molecules have been devised for using graphite in NIBs~\cite{Ri16jps,Yu17ea,Yu18pccp}, and also non-graphitic carbon materials such as soft and hard carbon have been developed, offering a significantly high capacity of $\sim$300 mA h g$^{-1}$ but with low Coulombic efficiency~\cite{Irisarri15jes}.
Titanium-based oxides are another promising candidate for the intercalation-type anodes because of their low voltage, low cost and non-toxicity~\cite{Zhai19jmca}.
Among them, sodium-titanate \ce{Na2Ti3O7} (NTO) with well-defined step-layered structure was found to exhibit the lowest electrode potential (0.3 V vs Na$^+$/Na) and high theoretical capacity (311 mA h g$^{-1}$~\cite{Wang13n}), demonstrated for the first time by Tarascon's group in 2011~\cite{Senguttuvan11cm}.
It was revealed that upon sodiation \ce{Na2Ti3O7} could transform into \ce{Na4Ti3O7} with the sluggish kinetics of \ce{Na+} ion insertion and the large lattice expansion (6\%), resulting in poor electrochemical performance.

Thereafter, NTO-based compounds have been extensively studied, aiming at improving the performance.
The synthesis was evolved from the simple solid-state reaction way~\cite{Senguttuvan11cm,Zarrabeitia16am,Pan13aem} to elaborate multistep reaction routes for specially designed phases~\cite{Xie17am,Ko17aam,Anwer17aam,Dong17jmca,Ni16aem,Fu16nl,Zarrabeitia16jps1,Marquez15aami,Zarrabeitia16jps2,Zou14ssi}.
Accordingly, the phase morphology was diversified from microsize particles~\cite{Senguttuvan11cm,Zou14ssi} to various nanoscale phases, including nanotubes~\cite{Ni16aem,Liu20aml,Yan17jac,Zhang14cc,Wang13n}, nanosheets~\cite{Ko17aam,Anwer17aam,Dong16afm}, nanowires~\cite{Zhou16ea}, nanofibers~\cite{Zou17cjc} and so on.
Such nanosizing was proved to effectively enhance the \ce{Na+} ion storage kinetics, increase the surface area and suppress the volume expansion.
Moreover, most nanoscale phases are in the form of complex with, for example, carbon~\cite{Zou17cjc,Ding17jps}, N-doped carbon~\cite{Xie17am}, graphene~\cite{Zeng18acie,Zhou16ea} and reduced graphene oxide~\cite{Bi18ea}, from which made flexible electrodes offered excellent rate capability and cycling stability.
Introducing other elements into NTO was reported to be another way for enhancing the electronic and ionic conductivity of electrode, such as lathanide-doping~\cite{Xia18cs}, Nb-doping~\cite{Chen17ea} and Ti$^{3+}$ self-doping~\cite{Song18jac,Rudola15ec}.
In spite of the excellent electrochemical performance, the nanophasing indeed greatly eliminated the plateau feature of NTO electrode in voltage profile.
Instead, the slope-like voltage profiles were observed, indicating their usability as battery-type (pseudocapacitance) anodes for NICs~\cite{Ko17aam,Dong17jmca,Dong16afm,Chen19aami,Qiu18jmca,Gao18jac,Dong15jmca}.
For instance, Dong and coworkers fabricated the quasi-solid-state NICs using 3D self-supported NTO nanoribbon array/graphene foam as the anode and graphene foam as the cathode, which delivered high specific energy of 70 Wh kg$^{-1}$, high specific power of 4000 W kg$^{-1}$, and superior cycling stability like capacitance retention 73\% over 5000 cycles~\cite{Dong17jmca}.

In comparison with such extensive experimental studies, only a few number of theoretical works to reveal the sodiation/desodiation mechanism has been reported so far~\cite{Wang13n,Pan13aem,Xu14cc,Avendano15jmca,Choe19pccp}.
Using first-principles simulations within the density functional theory (DFT) framework, Jiao and coworkers calculated formation energies of bulk Na$_x$Ti$_3$O$_7$ as increasing the Na content $x$, identifying the maximum value of $x$ to be 5.5 that gives the maximum theoretical capacity of 311 m Ah g$^{-1}$~\cite{Wang13n}.
Hu et al. calculated the band structure of NTO with a direct bandgap of 2.75 eV, which was smaller than their experimental value of 3.73 eV, and estimated the activation barriers for vacancy-mediated Na hopping along the layers and through the layers, indicating the former path surely to occur~\cite{Pan13aem}.
The theoretical electrode voltage profiles in reasonable agreement with experiment were calculated by Meng and coworkers~\cite{Xu14cc}.
However, a comprehensive study has not yet been provided on NTO and furthermore its vanadium-modified compounds, which were proposed to offer a better performance~\cite{Senguttuvan15usp}.

Herein we presented the electrochemical properties of NTO and its two different Ti/V exchange compounds, \ce{Na2Ti2VO7} (\ce{NTV1O}) and \ce{Na2TiV2O7} (\ce{NTV2O}), obtained by performing DFT simulations.
We identified the possible sites for Na insertion into NTO and pathways for Na ion migrations based on the bond valence sum (BVS) analysis.
As increasing the Na content, we calculated the formation energies of compounds and derived the convex hull plots, which were used to plot the electrode voltage profiles.
The activation energies for Na migrations along the paths identified through the BVS analysis were calculated to reveal the ionic conductivity and Na storage mechanism.
We calculated the electronic density of states in these compounds, providing an insight into their electronic transport.
Systematic comparison between NTO and NTVO compounds was given, emphasizing the positive role of Ti/V exchange in improving the electrode performance.

\section{Methods}
\subsection{Structural models}
\begin{figure*}[!th]
\centering
\includegraphics[clip=true,scale=0.13]{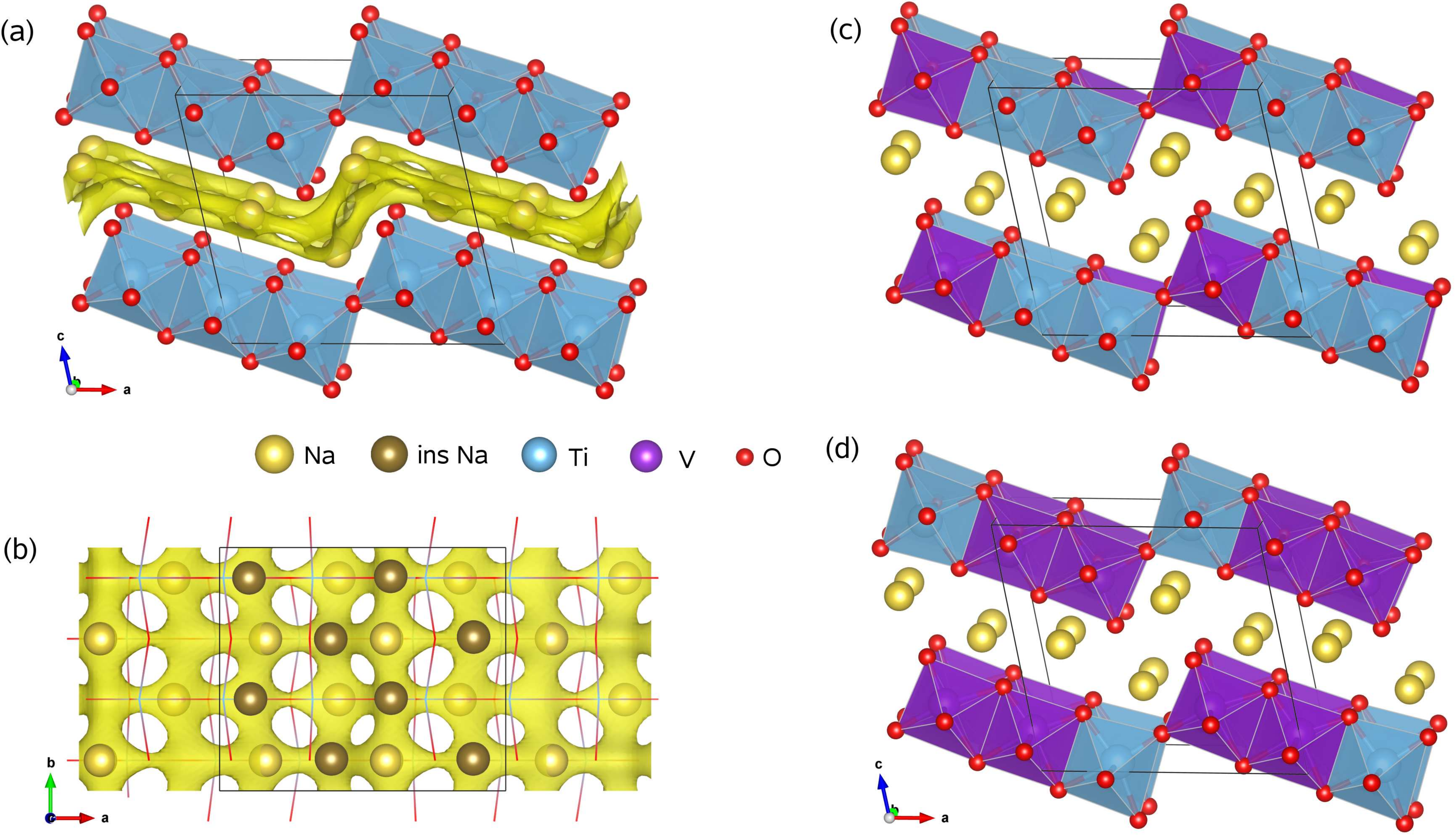}
\caption{Optimized structure of supercell containing 4 formula units for bulk \ce{Na2Ti3O7} in (a) perspective and (b) top views. Yellow-colored isosurface plot of bond valence sum at the value of 3 is shown to indicate the probable positions for additional inserting Na (ins Na) ions and their diffusion paths. Supercells for bulk (c) \ce{Na2Ti2VO7} and (d) \ce{Na2TiV2O7}, selected as the lowest energy configurations. Black solid lines indicate the lattice of supercell.}
\label{fig1}
\end{figure*}

The sodium titanate \ce{Na2Ti3O7} is known to crystallize in step-layered structure with a monoclinic space group $P2_1/m$, in which each step consists of three edge-sharing \ce{TiO6} octahedra connected with the others by vertex-sharing~\cite{Andersson61ac}.
The Na ions are located in the interlayer space with two different crystallographic coordinations (7 and 9).
By doubling the unit cell (2 formula units, 24 atoms) in (010) direction, we constructed a supercell (4 formula units, 48 atoms) to investigate Na insertion and diffusion, as shown in Fig.~\ref{fig1}(a).
It should be noted that the size of the supercell is similar to those of other NIB cathode materials in our previous works~\cite{yucj17pra,Ri18jps}, being expected to give reliable result for electrochemical properties of NTO.

In the present form of NTO, Ti is in +4 oxidation state, and upon insertion of additional \ce{Na+} ions, its oxidation state can change to +3, allowing NTO to act as electrode active material with \ce{Ti^{4+}}/\ce{Ti^{3+}} redox couple.
Therefore, it is necessary to identify probable positions for the additional \ce{Na+} ions, which can be carried out by plotting the BVS isosurface of NTO~\cite{Adam01acb,Wong17pccp}.
We calculated BVS for the Na$-$O bond in NTO using the formula, $B(\mathbi{r})=\sum_i\exp[(1.803-R_i(\mathbi{r})/0.37]$ ($R(\mathbi{r})=|\mathbi{r}-\mathbi{R}_i|$, $\mathbi{R}_i$ is the position vector of the $i$-th atom), and plotted the BVS isosurface at the value of 3, as shown in Fig.~\ref{fig1}(a) and (b).

In the supercell containing 4 formula units, the eight host Na ions are shown to occupy the most probable sites, and another 8 equivalent positions are found to be unoccupied in the interlayer space.
Accordingly, 8 Na ions in total can be inserted one by one into the interlayer space of supercell, forming Na$_{2+x}$\ce{Ti3O7} ($x=0-2$), while some Ti ions are reduced from +4 to +3 oxidation state.
To reduce all the \ce{Ti^{4+}} ions to \ce{Ti^{3+}}, 4 Na ions should be more inserted into the supercell, resulting in the formation of \ce{Na5Ti3O7}.
To search these positions, we again drew the BVS isosurface plot at higher value of 14, identifying three sites more in the gap spaces between the neighboring up and down steps of layer (see Fig. S1 in Supporting Information).
However, we disregarded these sites because they were found at much higher value of BVS (compared with the ideal value of 1) and moreover the crystalline phase formed by filling these sites with Na ions could be severely deviated from the original phase of NTO, leading to large volume expansion and thus short cycling lifetime~\cite{Wang13n}.
In fact, the \ce{Na4Ti3O7} compound rather than \ce{Na5Ti3O7} was found to be the final production in the sodium storage process in the most experimental works~\cite{Senguttuvan11cm}.

To investigate the influence of Ti exchange with V on sodium storage performance, we considered two different exchange models, \ce{Na_{$x$}Ti2VO7} and \ce{Na_{$x$}TiV2O7}.
Since vanadium has higher oxidation state of +5 than Ti, the Na content $x$ can be below 2, indicating that the host Na ions can be removed from \ce{Na2Ti_{3-$x$}V_{$x$}O7} ($x$ = 1 and 2).
In these Ti/V exchange compounds, the lowest values of Na content reduce to 1 and 0 for \ce{NTV1O} and \ce{NTV2O}, while the highest values are equal to 5.
With these considerations, we can write the reduction processes occurred on NTO, \ce{NTV1O} and \ce{NTV2O} upon insertion of Na ions into their lowest Na content compounds as follows,
%
\begin{subequations}
\begin{gather}
\ce{Na2Ti}^{+4}_3\ce{O}_7 + 3\ce{Na} \rightarrow \ce{Na5Ti}^{+3}_3\ce{O}_7, \label{eq1a}\\
\ce{NaTi}^{+4}_2\ce{V}^{+5}\ce{O}_7 + 4\ce{Na} \rightarrow \ce{Na5Ti}^{+3}_2\ce{V}^{+3}\ce{O}_7, \label{eq1b}\\
\ce{Ti}^{+4}\ce{V}^{+5}_2\ce{O}_7 + 5\ce{Na} \rightarrow \ce{Na5Ti}^{+3}\ce{V}^{+3}_2\ce{O}_7. \label{eq1c}
\end{gather}
\end{subequations}
%
Then, we can expect an enhancement of electrode specific capacity by Ti/V exchange in NTO, as discussed below.

It is important to provide concrete evidence of formability as experimental work on synthesis of the NTVO compounds could not yet found.
As a first check for formability, we calculate cohesive energy $E_c$ using the following equation,
\begin{equation}
E_c=E_{\ce{Na2Ti_aV_bO7}}-(2E_{\ce{Na}}^i+aE_{\ce{Ti}}^i+bE_{\ce{V}}^i+7E_{\ce{O}}^i) \label{eq_ec}
\end{equation}
where $E_{\ce{Na2Ti_aV_bO7}}$ is the total energy per formula unit of \ce{Na_2Ti_{$a$}V_{$b$}O7} bulk and $E^i_X$ is the total energy of isolate atom $X$.
We also calculate the formation enthalpy of the compounds from binary metal oxides including \ce{Na2O} in cubic phase (space group $FM\bar{3}M$), anatase-type \ce{TiO2} and \ce{V2O5} in orthorhombic phase ($PMMN$).
For the following chemical reactions for synthesis,
\begin{equation}
\begin{gathered}
\ce{Na2O}+3\ce{TiO2}\Rightarrow\ce{Na2Ti3O7}, \\
2\ce{Na2O}+4\ce{TiO2}+\ce{V2O5}\Rightarrow 2\ce{Na2Ti2VO7}+\tfrac{1}{2}\ce{O2}, \\
\ce{Na2O}+\ce{TiO2}+\ce{V2O5}\Rightarrow \ce{Na2TiV2O7}+\tfrac{1}{2}\ce{O2} \\
\end{gathered}
\end{equation}
the formation enthalpy $E_f^{\text{b}}$ can be calculated as follows,
\begin{equation}
\begin{gathered}
E_f^{\text{b}}=E_{\ce{Na2Ti3O7}}-(E_{\ce{Na2O}}+3E_{\ce{TiO2}}), \\
E_f^{\text{b}}=\frac{1}{2}\left[2E_{\ce{Na2Ti2VO7}}+\tfrac{1}{2}E_{\ce{O2}}-(2E_{\ce{Na2O}}+4E_{\ce{TiO2}}+E_{\ce{V2O5}})\right], \label{eq_efb} \\
E_f^{\text{b}}=E_{\ce{Na2TiV2O7}}+\tfrac{1}{2}E_{\ce{O2}}-(E_{\ce{Na2O}}+E_{\ce{TiO2}}+E_{\ce{V2O5}})
\end{gathered}
\end{equation}
where $E_{\ce{O2}}$ is the total energy of isolated \ce{O2} molecule that can be simulated by supercell with lattice constant of 15 \AA.
On the other hand, the synthesis was suggested to be performed by using ternary sodium carbonate \ce{Na2CO3} instead of binary sodium oxide \ce{Na2O}~\cite{Senguttuvan15usp} as,
\begin{equation}
\begin{gathered}
\ce{Na2CO3}+3\ce{TiO2}\Rightarrow\ce{Na2Ti3O7}+\ce{CO2}, \\
2\ce{Na2CO3}+4\ce{TiO2}+\ce{V2O5}\Rightarrow 2\ce{Na2Ti2VO7}+2\ce{CO2}+\tfrac{1}{2}\ce{O2}, \\
\ce{Na2O}+\ce{TiO2}+\ce{V2O5}\Rightarrow \ce{Na2TiV2O7}+\ce{CO2}+\tfrac{1}{2}\ce{O2} \\
\end{gathered}
\end{equation}
for which the formation enthalpy $E_f^{\text{t}}$ can be calculated as follows,
\begin{equation}
\begin{gathered}
E_f^{\text{t}}=E_{\ce{Na2Ti3O7}}+E_{\ce{CO2}}-(E_{\ce{Na2CO3}}+3E_{\ce{TiO2}}), \\
E_f^{\text{t}}=\frac{1}{2}\left[2E_{\ce{Na2Ti2VO7}}+2E_{\ce{CO2}}+\tfrac{1}{2}E_{\ce{O2}}-(2E_{\ce{Na2CO3}}+4E_{\ce{TiO2}}+E_{\ce{V2O5}})\right], \label{eq_eft} \\
E_f^{\text{t}}=E_{\ce{Na2TiV2O7}}+E_{\ce{CO2}}+\tfrac{1}{2}E_{\ce{O2}}-(E_{\ce{Na2CO3}}+E_{\ce{TiO2}}+E_{\ce{V2O5}})
\end{gathered}
\end{equation}
For the cases of bulk, we can assume that the enthalpic difference term $P\Delta V$ and the entropic term $T\Delta S$ at room temperature are negligible due to being 3 or 4 order smaller than the internal energy difference.
Meanwhile, for the cases of \ce{O2} and \ce{CO2} gas we should consider both terms which can be taken from experimental data~\cite{Lide05crc}.

\begin{table*}[!th]
\small
\caption{Optimized lattice parameters and volume of supercell with 4 formula units, cohesive energy $E_c$ (eV per atom), and formation enthalpies at $T=298.15$ K from binary oxides $E^b_f$ and from ternary compound  $E^t_f$ (eV per formula unit) of the electrode materials}
\label{tabl1}
\begin{tabular}{lcccccccccc}
\hline
Compound & $a$ (\AA) & $b$ (\AA) & $c$ (\AA) &$\alpha$ ($^\circ$) & $\beta$ ($^\circ$) & $\gamma$ ($^\circ$) & Vol (\AA$^3$) & $E_c$ & $E^b_f$ & $E^t_f$ \\
\hline
\ce{Na2Ti3O7} & 9.040 & 7.519 & 8.475 & 90.00 & 102.04 & 90.00 & 563.36 & $-5.57$ & $-2.25$ & $-0.10$ \\
\hspace{18pt}Exp$^a$ & 9.126 & 7.599 & 8.563 & 90.00 & 101.59 & 90.00 & & & & \\
\ce{Na2(Ti2V)O7} & 8.948 & 7.483 & 8.433 & 89.99 & 102.07 & 90.00 & 552.15 & $-5.55$ & $-2.21$ & $-0.05$ \\
\ce{Na2(TiV2)O7} & 8.885 & 7.465 & 8.393 & 90.00 & 101.65 & 90.00 & 545.27 & $-5.53$ & $-2.19$ & $-0.04$ \\
\hline
\end{tabular} \\
$^a$ X-ray diffraction data~\cite{Pan13aem}.
\end{table*}

\subsection{Computational details}
All the DFT calculations have been carried out by applying the pseudopotential plane wave method as implemented in the {\footnotesize QUANTUM ESPRESSO} package~\cite{QE09jpcm}.
To describe the interaction between the valence electrons and ionic cores, we have used the ultrasoft pseudopotentials established in the package, where the valence electron configurations were set to Na:$2s^22p^63s^1$, Ti:$3s^23p^63d^{2}4s^2$, V:$3s^23p^63d^{3}4s^2$, and O:$2s^22p^4$.
The Perdew-Burke-Ernzerhof (PBE) formulation~\cite{PBE96prl} within generalized gradient approximation (GGA) was adopted for the exchange-correlation (XC) interaction between the valence electrons.
The weak van der Waals (vdW) interaction between the layers was considered through the semi-empirical Grimme's function (DFT-D2)~\cite{Grimme06jcc}.
We have also applied the GGA + $U$ approach in the simplified version of Cococcioni and de Gironcoli~\cite{Cococcioni05prb} to take into account the effect of localized $3d$ electrons of the transition metal cations.
The effective Hubbard parameter $U_{\text{eff}}=3$ eV was given to Ti and V atoms, consistent with the previous work~\cite{Xu14cc}.
Structural optimizations of the supercells were carried out with cutoff energies of 50 Ry for wave function and 500 Ry for electron density and a $(2\times 2\times 2)$ $k$-point mesh until the force on each atom and the pressure of crystalline lattice converged to $5\times 10^{-4}$ Ry Bohr$^{-1}$ and 0.05 Kbar respectively.
To calculate the density of states (DOS), a denser $k$-point mesh of $(6\times 6\times 6)$ was adopted with consideration of spin-polarization effect, where ferromagnetic ordering was applied.

The activation energies for Na ion diffusion according to the paths predicted by BVS in the supercells were calculated by applying the climbing image nudged elastic band (NEB) method~\cite{NEB00jcp}.
During the NEB simulation, the supercell sizes were fixed at the optimized ones, and all the atoms were allowed to relax until the force converged to 0.02 eV \AA$^{-1}$.
The number of NEB image points was set to 7 or 13 according to the route of Na ion migration.
Visualization of crystalline lattice, and volumetric data of BVS and electronic density was carried out by using the VESTA code~\cite{VESTA11jac}

\section{Results and discussion}
\subsection{Structural variation}
Firstly, we determined the lattice parameters of NTO unit cell with 2 formula units in monoclinic phase by using different XC functionals.
With the PBE + DFT-D2 functional, they were calculated to be $a=9.092$ \AA, $b=3.786$ \AA, $c=8.510$ \AA, $\beta=101.88^{\circ}$, which were slightly underestimated compared with the experimental values within a relative error of $-$0.5\%.
Such underestimation is due to the inclusion of vdW dispersion energy, since the pure PBE functional gave the slight overestimation within the well-known trend of the GGA functional.

Then, the supercell with 4 formula units was constructed by doubling the optimized unit cell in the (010) direction, which was again optimized.
The polyhedral view of its crystal structure is shown in Fig.~\ref{fig1} (a) and (b), together with the isosurface plot of BVS.
To construct the supercells for \ce{NTV1O} and \ce{NTV2O} crystalline solids, 4 and 8 Ti ions were exchanged with V ions in the supercell.
By considering the crystallographic equivalent positions, 15 different configurations for Ti/V exchange were considered to be possible and the corresponding supercells were also optimized.
Among them, the lowest energy configurations were selected, and their optimized structures are shown in Fig.~\ref{fig1}(c) and (d) for \ce{NTV1O} and \ce{NTV2O}, respectively.
Upon such replacement of some Ti ions with V ions, the lattice parameters of supercells were found to decrease as increasing the content of V ions, as listed in Table~\ref{tabl1}.
Accordingly, the volumes of the supercells were also observed to decrease by 98.0\% and 96.8\% for \ce{NTV1O} and \ce{NTV2O} respectively, compared with NTO.
These are possibly due to an enhancement of interaction between the transition metal cations and oxygen anions as reflected by the average M$-$O bond length within the \ce{MO6} octahedra (1.98 \AA~in \ce{TiO6} vs 1.93 \AA~in \ce{VO6}).
In Table~\ref{tabl1} we also presented the cohesive energy, and binary and ternary formation enthalpies calculated by applying Eq.~\ref{eq_ec}, and Eqs.~\ref{eq_efb} and \ref{eq_eft} respectively.
The cohesive energy and formation enthalpies were calculated to be negative for both \ce{NTV1O} and \ce{NTV2O} as well as NTO, indicating that they are thermodynamically stable and can be synthesized securely.
As increasing the amount of V, the energy was found to slightly decrease in magnitude, implying the slackening of stability. 

During the charge/discharge process in NIBs, Na ions should be inserted/deserted into/from the intercalation-type anode material, causing a volume change or even multi-step phase transition.
To simulate the charge process, we constructed the supercells for NTO, \ce{NTV1O} and \ce{NTV2O} with increasing number of Na ions from 1 to 8, which were placed on the sites identified by BVS analysis as discussed above.
It should be emphasized that the final productions are \ce{Na4M3O7} based on the BVS analysis in this work, although \ce{Na5M3O7} is theoretically possible as proved in Eqs.~\ref{eq1a}$-$\ref{eq1c}.
Meanwhile, we considered the supercells with remove of host Na ions up to 4 and 8 for the cases of \ce{NTV1O} and \ce{NTV2O} respectively. 
There were many configurations for Na insertion and remove, which could be reduced by considering the crystalline symmetry.
After picking out the lowest energy configuration of each intermediate phase with structural optimization, we measured the lattice parameters and volume as increasing the number of Na ions.

\begin{figure}[!th]
\centering
\includegraphics[clip=true,scale=0.5]{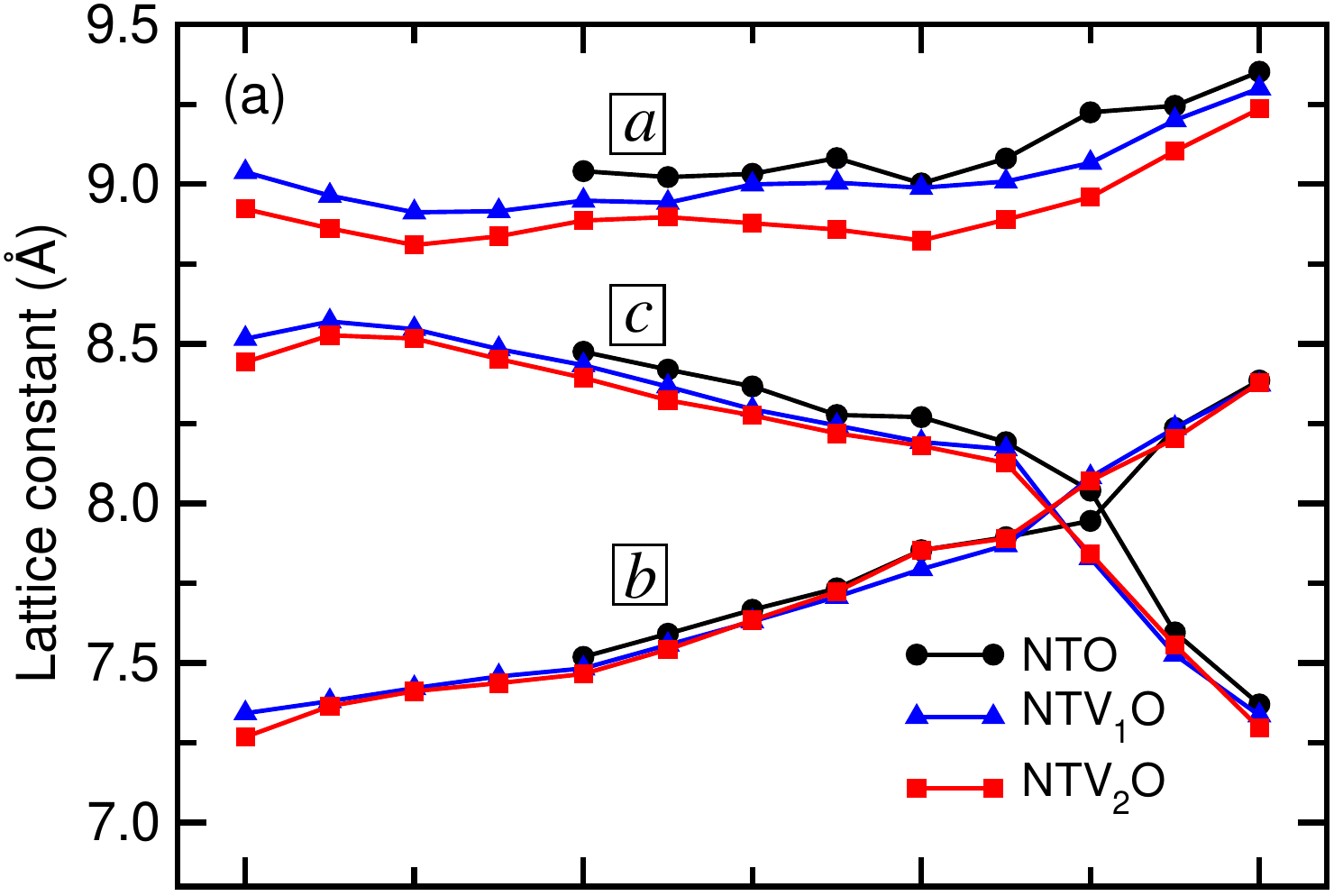}~~~~~~~~\\
\includegraphics[clip=true,scale=0.5]{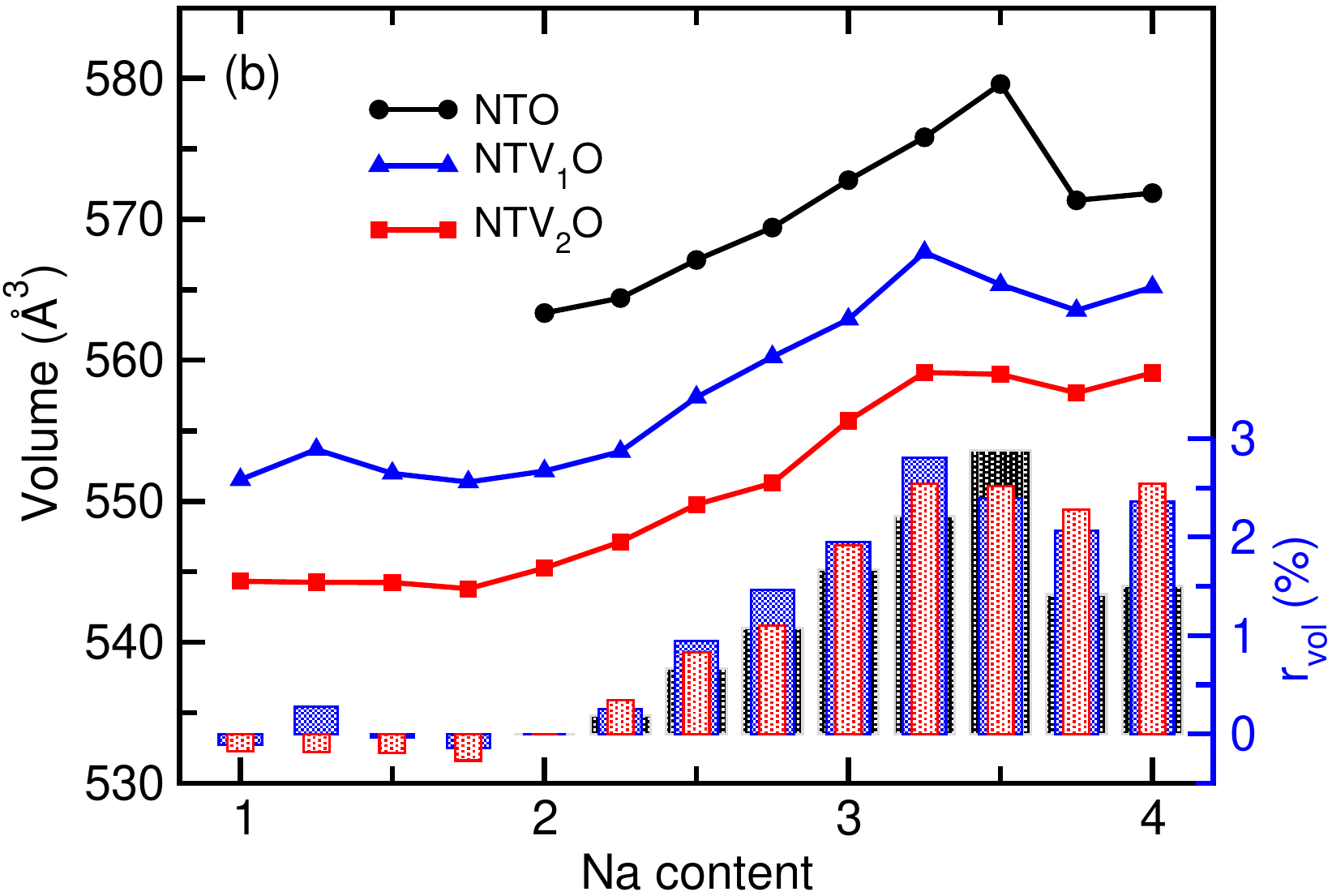}
\caption{Variation of (a) lattice constants ($a, b, c$) and (b) volume of supercell with 4 formula units as increasing the Na content $x$ in \ce{Na_{$x$}M3O7} (M = Ti, Ti$_{2/3}$V$_{1/3}$, Ti$_{1/3}$V$_{2/3}$), where the relative volume expansion rate $r_{\text{vol}}=(V_x-V_{x=2})/V_{x=2}\times 100$(\%) is shown by rectangular bar.}
\label{fig2}
\end{figure}
In Fig.~\ref{fig2}, we show the calculated lattice constants and volume as a function of Na content $x$ in Na$_x$\ce{M3O7} compounds.
For the case of M = Ti (NTO), the Na content varies from $x=2$ to 4, while for the cases of M = Ti$_{2/3}$V$_{1/3}$ and Ti$_{1/3}$V$_{2/3}$ it starts from $x=1$.
It should be noted that for the case of \ce{NTV2O}, the supercell volume was found to abruptly decrease when removing the host Na ions by from 5 ($x=1$) to 8 ($x=0$) (see Fig. S2 in Supporting Information), implying that Na$_x$\ce{TiV2O7} with $x\in (0, 1)$ can not be expected to properly work as electrode active material.
As increasing the Na content $x$, the lattice constants $b$ are shown in Fig.~\ref{fig2}(a) to gradually increase and $a$ to increase in more or less wavy way, whereas the interlayer lattice constants $c$ to decrease gradually first and rapidly after $x=3.25$.
This indicates that the insertion of Na ions into the interlayer space enhances the interaction between the layers through the attractive Na$-$O bonding.
In most cases of Na content, the lattice constants became smaller as increasing V content, indicating that Ti/V exchange in NTO leads to enhancement of M$-$O interaction.
On the other hand, the supercell volumes were found to gradually increase from $x=1$ for NTVO and $x=2$ for NTO to $x=3.25$ and decrease after $x=3.25$, as can be seen in Fig.~\ref{fig2}(b).
The relative volume expansion rates, $r_{\text{vol}}=(V_x-V_{x=2})/V_{x=2}\times 100$(\%), were found to be under 3\%, which is quite small compared with other electrode materials, indicating their superior cycling stabilities.
For all the three cases of electrode materials, similar variation tendencies and moreover no phase transitions were observed.
It is worth noting that from $x=1$ to $x=2$ for Ti/V exchange models the volume changes are almost negligible.

\subsection{Electrode potential}
We estimated the relative stability of NTO-based compounds upon insertion and extraction of Na ions by calculating the formation energy of intermediate \ce{Na_{$n$+$x$}M3O7} compound with respect to the two end compounds.
As mentioned above, the starting compounds in this work were arranged to be \ce{Na2M3O7} for all the cases and \ce{NaM3O7} only for NTVO cases, correspondingly indicating $n$ to be 2 and 1, while the final compounds were consistently \ce{Na4M3O7}.
Then, the formation energies can be written in two ways, as follows,
\begin{subequations}
\begin{gather}
E_{\text{f}}=E_{\ce{Na_{2+x}M3O7}}-\frac{1}{2}\left[xE_{\ce{Na4M3O7}}+(2-x)E_{\ce{Na2M3O7}} \right], \label{eq2a}\\
E_{\text{f}}=E_{\ce{Na_{1+x}M3O7}}-\frac{1}{3}\left[xE_{\ce{Na4M3O7}}+(3-x)E_{\ce{NaM3O7}} \right], \label{eq2b}
\end{gather}
\end{subequations}
where $E_{\ce{comp}}$ is the DFT total energy of the compound.

\begin{figure}[!t]
\centering
\includegraphics[clip=true,scale=0.5]{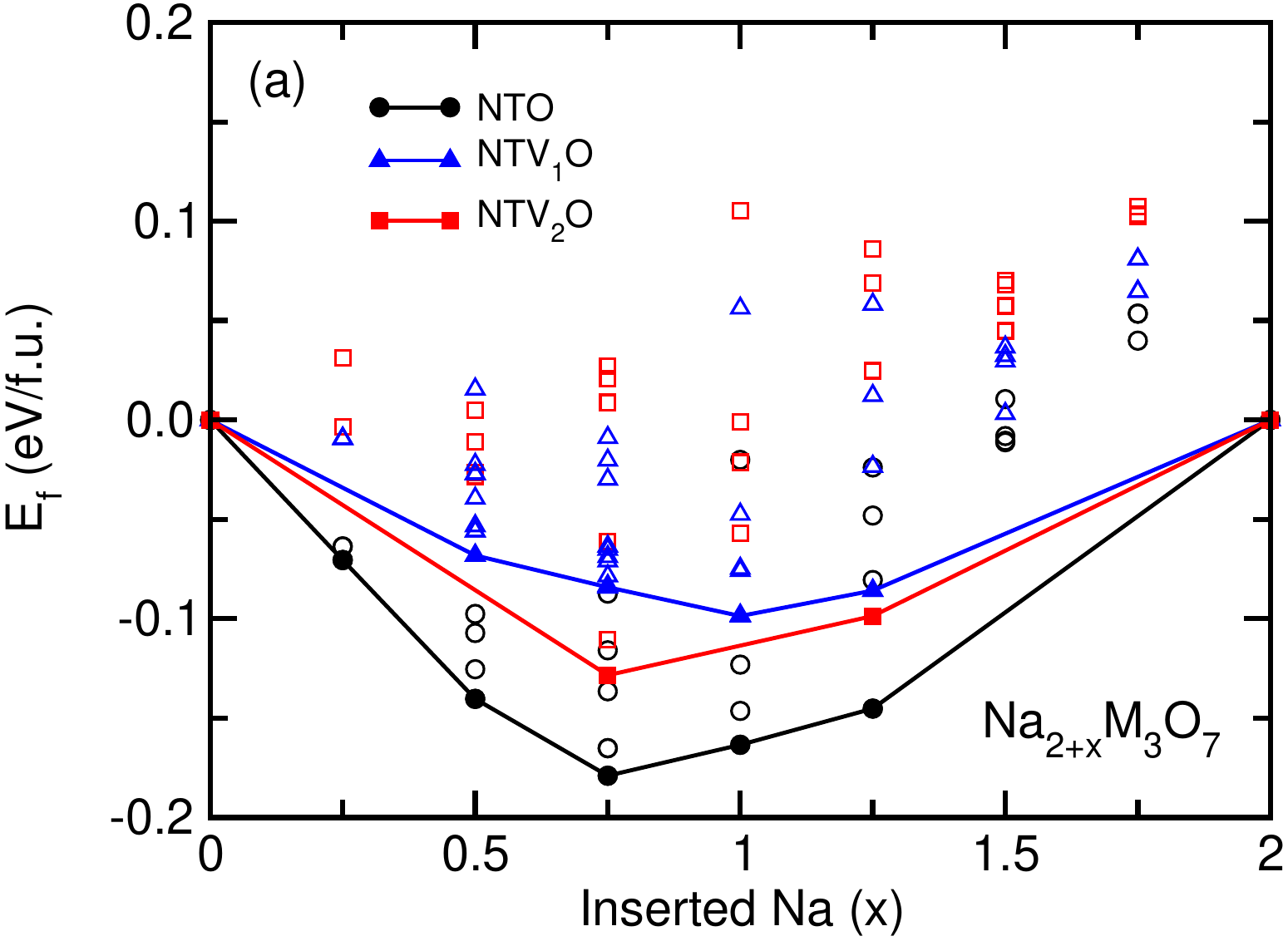} \vspace{2.5pt}\\
\includegraphics[clip=true,scale=0.5]{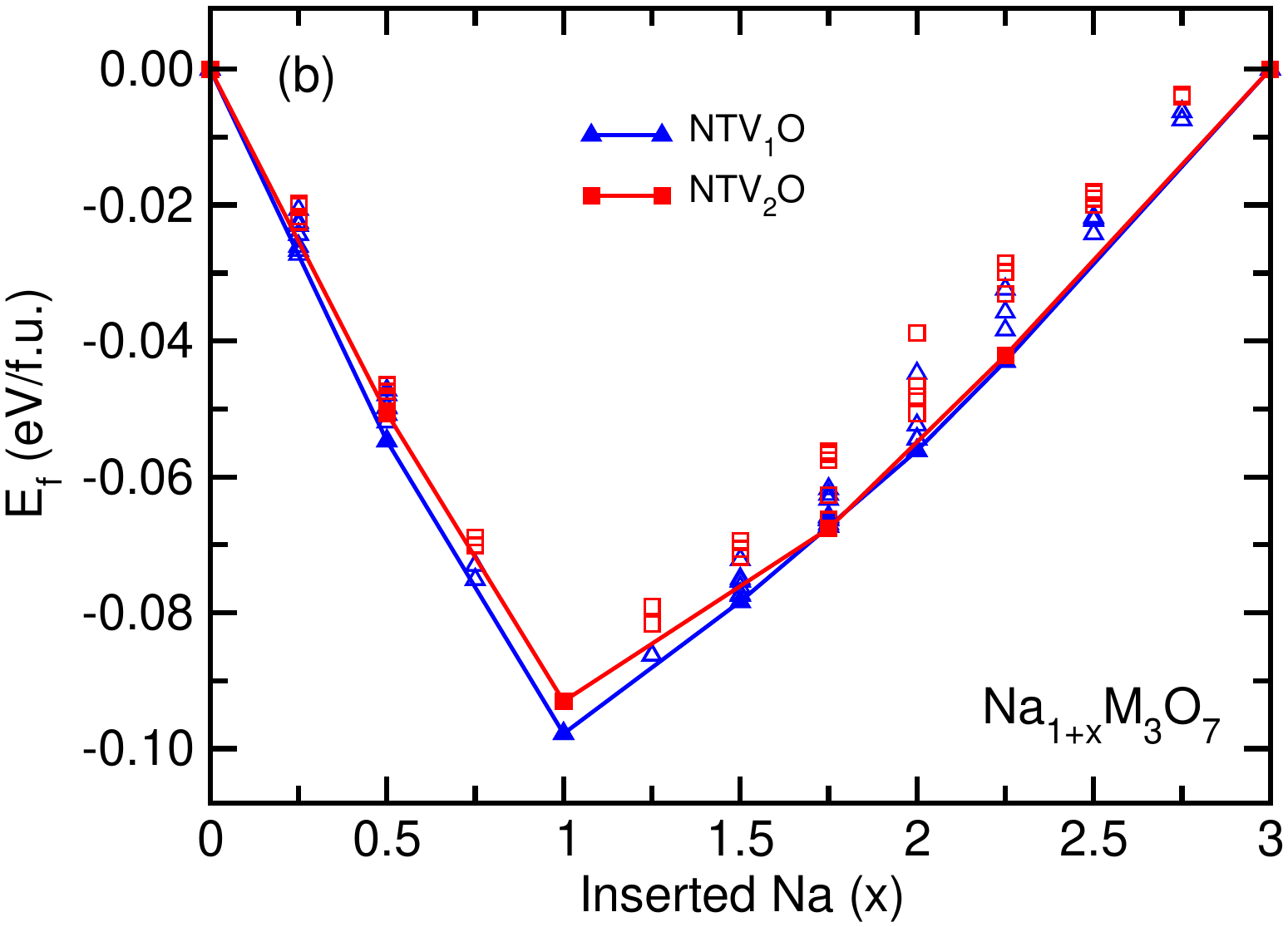}
\caption{Convex hull plot of formation energy per formula unit (f.u.) of \ce{Na2Ti3O7} (NTO), \ce{Na2Ti2VO7} (\ce{NTV1O}) and \ce{Na2TiV2O7} (\ce{NTV2O}), calculated with PBE + vdW (DFT-D) functional. The curves are obtained by using the starting compound as (a) \ce{Na2M3O7} for all the three cases of compounds and (b) \ce{NaM3O7} only for NTVO compounds.}
\label{fig3}
\end{figure}
Figure~\ref{fig3} shows the formation energies as a function of Na content, calculated by adopting the starting compound as \ce{Na2M3O7} for the three different kinds of models, and as \ce{NaM3O7} only for Ti/V exchange models.
In Fig.~\ref{fig3}(a) for \ce{Na_{2+$x$}M3O7} models, the formation energies of the intermediated phases with $x\leq 1.25$ are shown to be negative, indicating their safe formation, whereas the phases with $x=1.5$ and $x=1.75$ can not be said to be naturally formed due to their positive formation energies.
This is consistent with the abrupt decrease of the interlayer lattice constant $c$ and volume decrease with these Na contents, shown in Fig.~\ref{fig2}.
In these cases, the lowest energy configurations were found at $x=0.75$ for NTO and \ce{NTV2O}, while at $x=1$ for \ce{NTV1O} model.
When compared NTVO models with NTO, the formation energies are in overall higher, indicating that Ti/V exchange enhances the interactions between the cations and anions.

On the other hand, the formation energies, calculated by applying Eq.~\ref{eq2b} and using the two end compounds of \ce{NaM3O7} and \ce{Na4M3O7} for NTVO models, were found to be negative for the whole range of inserted Na content $x\in (0, 3)$, as shown in Fig.~\ref{fig3}(b).
In these cases, the intermediate phases with the lowest formation energy were identically found at $x=1$, {\it i.e.} \ce{Na2M3O7}, which is agreed well with the fact that it is the actual starting compound.
When compared between two different Ti/V exchange models of \ce{NTV1O} and \ce{NTV2O}, the former compounds were always shown to be more readily formed due to their slight lower formation energies.
We also calculated the formation energies of the intermediate phases with respect to the same starting compounds and Na metal in body-centered cubic (bcc) phase, and Na binding energies by using Na reservoir in gas state (see Fig. S3 in Supporting Information).
As the Na and V contents increase, the formation and Na binding energies were found to decrease in agreement with the above discussion.
The optimized structures for \ce{Na_{2+$x$}Ti3O7} ($x\in (0,2)$), \ce{Na_{1+$x$}Ti2VO7} ($x\in (0,3)$) and \ce{Na_{$x$}Ti2VO7} ($x\in (0,4)$) were shown in Fig. S4-S6 in Supporting Information.

In order to shed light on the enhancement of electrode performance by Ti/V exchange, we estimated the electrode potential $V$ as a function of specific capacity.
The step discharge voltage between the adjacent Na contents, identified as placed on the convex hull curves in Fig.~\ref{fig3}, with respect to the Na/\ce{Na+} counter electrode can be calculated as follows,
\begin{equation}
V=-\frac{E_{\ce{Na}_{x_j}\ce{M3O7}}-E_{\ce{Na}_{x_i}\ce{M3O7}}+(x_j-x_i)E_{\ce{Na}_{\text{bcc}}}}{(x_j-x_i)e}, \label{eq3}\\
\end{equation}
where $e$ is the elementary charge.

\begin{figure}[!t]
\centering
\includegraphics[clip=true,scale=0.5]{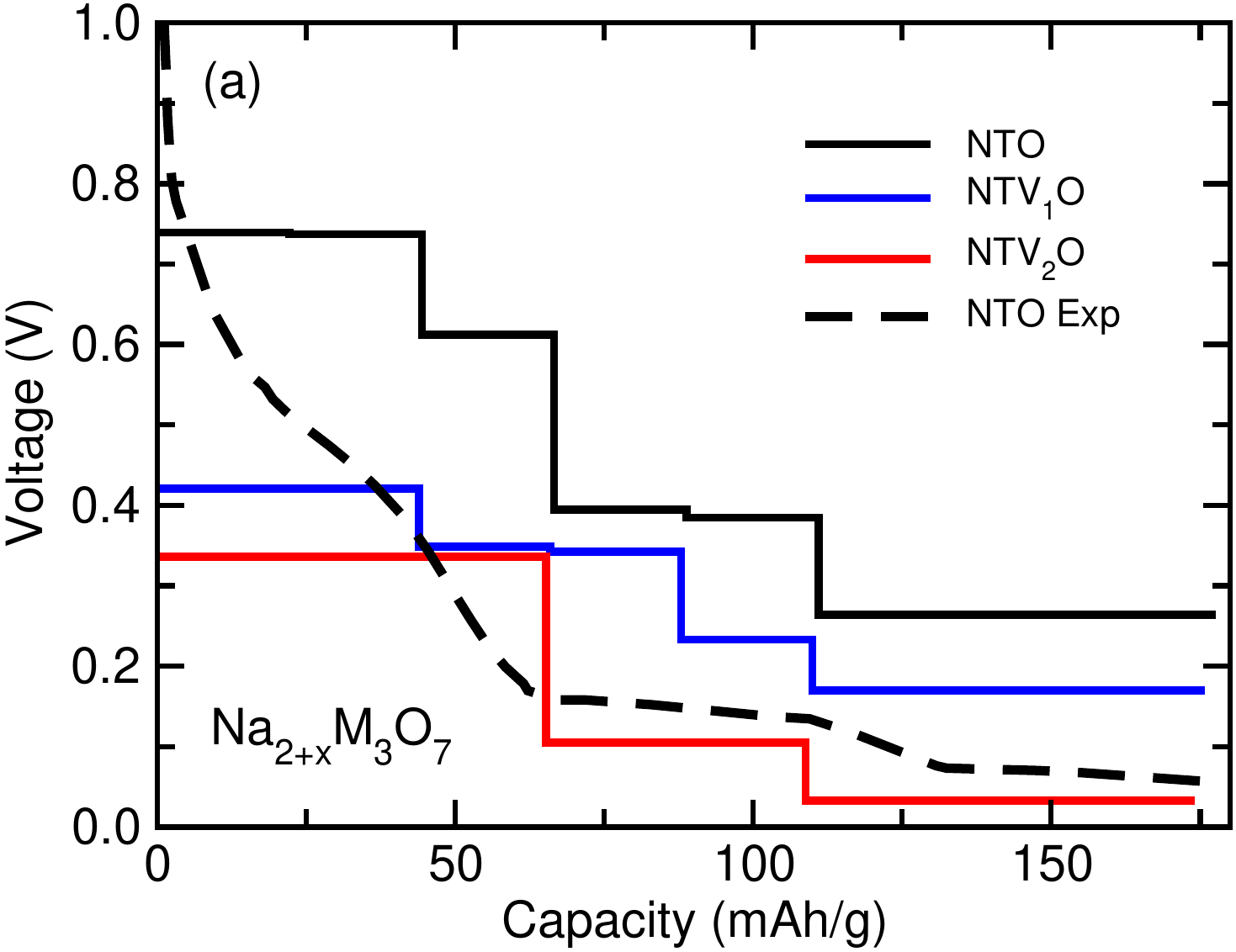} \\
\includegraphics[clip=true,scale=0.5]{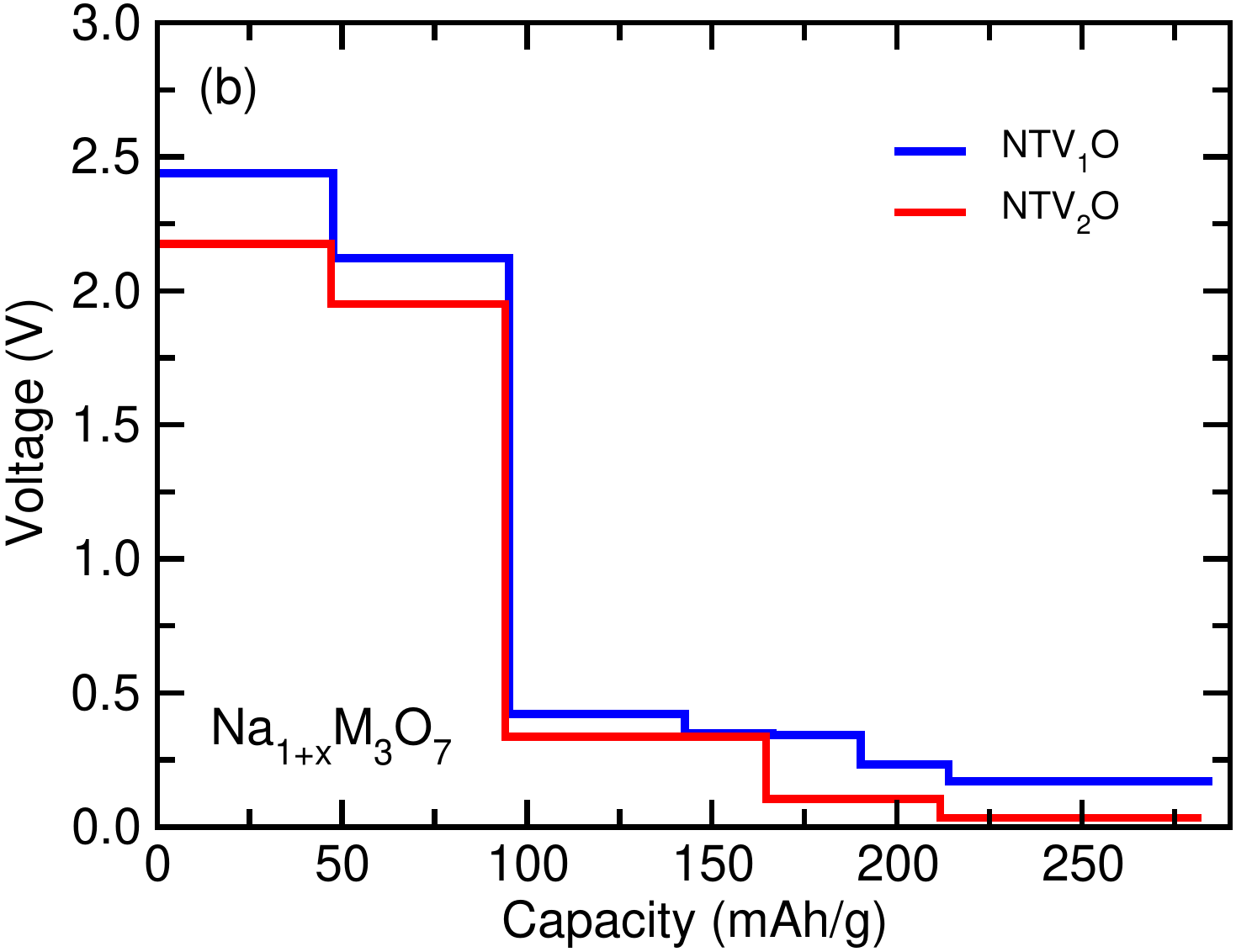}
\caption{Electrode potential as a function of specific capacity of of NTO, \ce{NTV1O} and \ce{NTV2O}, calculated by using the DFT total energies of the compounds identified as placed on the convex hull line (Fig.~\ref{fig3}). (a) \ce{Na_{2+$x$}M3O7} ($x\in (0, 2)$) for all the models, where the experimental curve for NTO is from Ref.~\cite{Marquez15aami}, and (b) \ce{Na_{1+x}M3O7} ($x\in (0, 3)$) for NTVO models.}
\label{fig4}
\end{figure}
Due to the two different working ways of electrodes distinguished by selecting the base compound, which is \ce{Na2M3O7} or \ce{NaM3O7}, we plotted also two different step voltage curves, as shown in Fig.~\ref{fig4}.
With \ce{Na2M3O7} as the starting compound, the maximal specific capacities were estimated to be 177.6, 175.8 and 174.1 mAh g$^{-1}$ in \ce{Na4Ti_{3-$x$}V_{$x$}O7} for $x$ =0, 1 and 2 respectively.
As shown in Fig.~\ref{fig4}(a), such lowering in capacity by Ti/V cation exchange is negligible but voltage lowering is clearly distinctive.
Together with the calculated step voltages below 1 V, this indicates that these compounds can work as anodic materials for NIB and Ti/V exchange can increase the full-cell voltage and energy density.
More interestingly, due to the higher oxidation state of \ce{V^{+5}}, \ce{NaTi_{3-$x$}V_{$x$}O7} ($x$ =1 and 2 in this work) can be thought as the base compound, resulting in a great increase of specific capacity.
As shown in Fig.~\ref{fig4}(b), the maximal capacities in these cases were determined to be 285.3 and 282.2 mAh g$^{-1}$ in \ce{Na4Ti_{3-$x$}V_{$x$}O7} for $x$ = 1 and 2, being much higher than those obtained by using \ce{Na2M3O7} as the base compound.

\subsection{Sodium-ion conductivity}
\begin{figure}[!b]
\centering
\includegraphics[clip=true,scale=0.12]{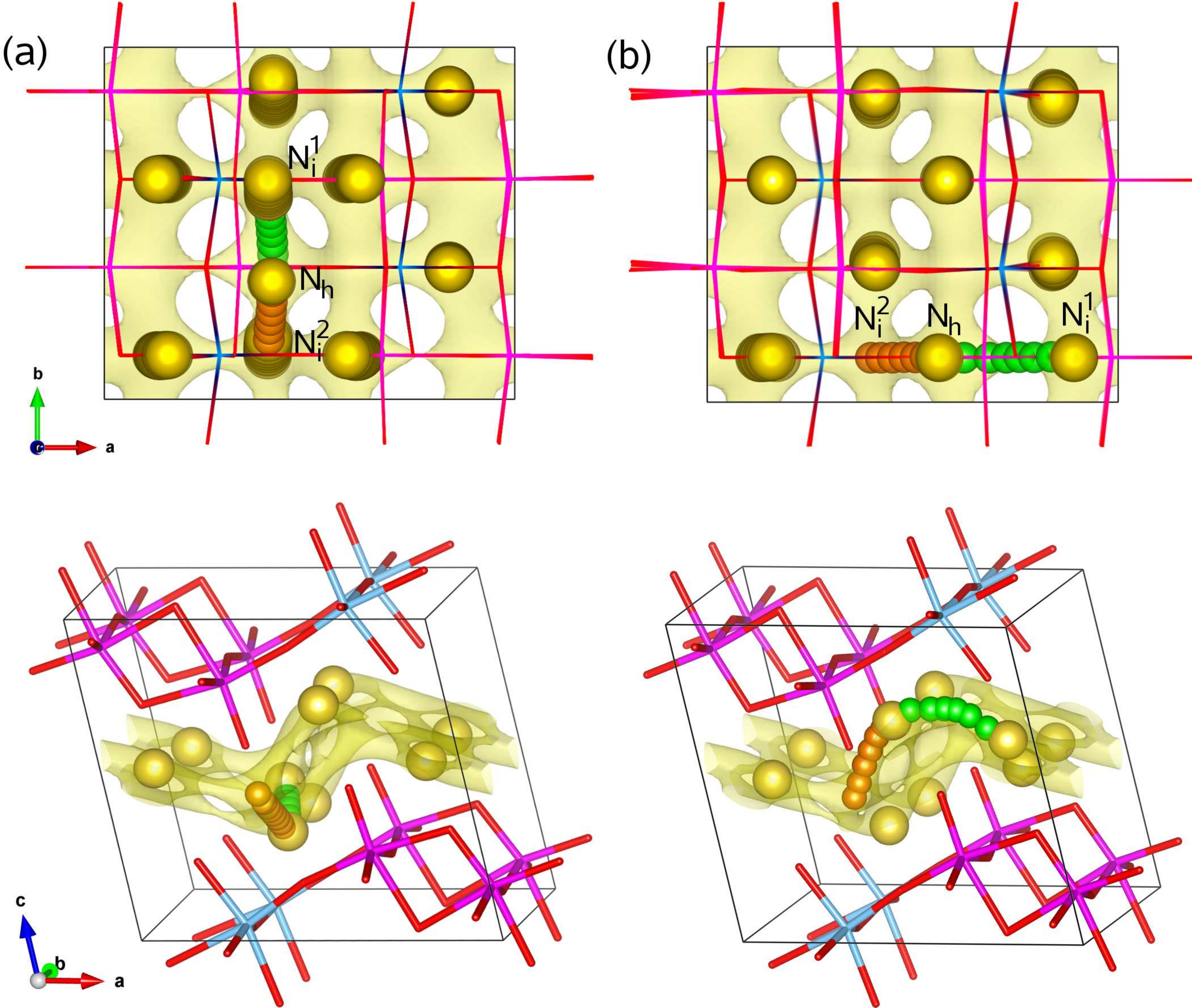} 
\caption{Pathways of Na ion migration along (a) (010) direction with a wavy-line character and (b) (100) direction with a step-line-step character, formed in the interlayer space and indicated by yellow-colored isosurface plot of BVS at the value of 3. Top and bottom panels show top and perspective views. Fixed and migrating Na atoms are denoted as yellow and green or orange balls, and Ti$-$O and V$-$O bonds as blue-red and purple-red sticks, respectively.}
\label{fig5}
\end{figure}
Ionic conductivity plays a decisive role in electrochemical performance of rechargeable batteries and capacitors.
To get an insight into ionic conductance, we determined the activation energies for vacancy-mediated Na ion migrations along the most plausible pathways, which were preliminary predicted by the BVS analysis.
As shown in Fig.~\ref{fig5}, the pathways can be formed in the interlayer space along the crystallographic (010) and (100) directions.
The (010) direction pathway can be characterized by a slightly wavy line without step, whereas along the (100) direction a step-line-step pathway can be formed.
In accordance with the exclusion of possible Na sites in the gap space between up and down steps, we did not consider the migration through the layer, for which the activation energies were determined to be much higher ($\sim$1.75 eV) than those for migrations along the layer with first-principles NEB calculation~\cite{Pan13aem}.

\begin{figure}[!b]
\centering
\includegraphics[clip=true,scale=0.5]{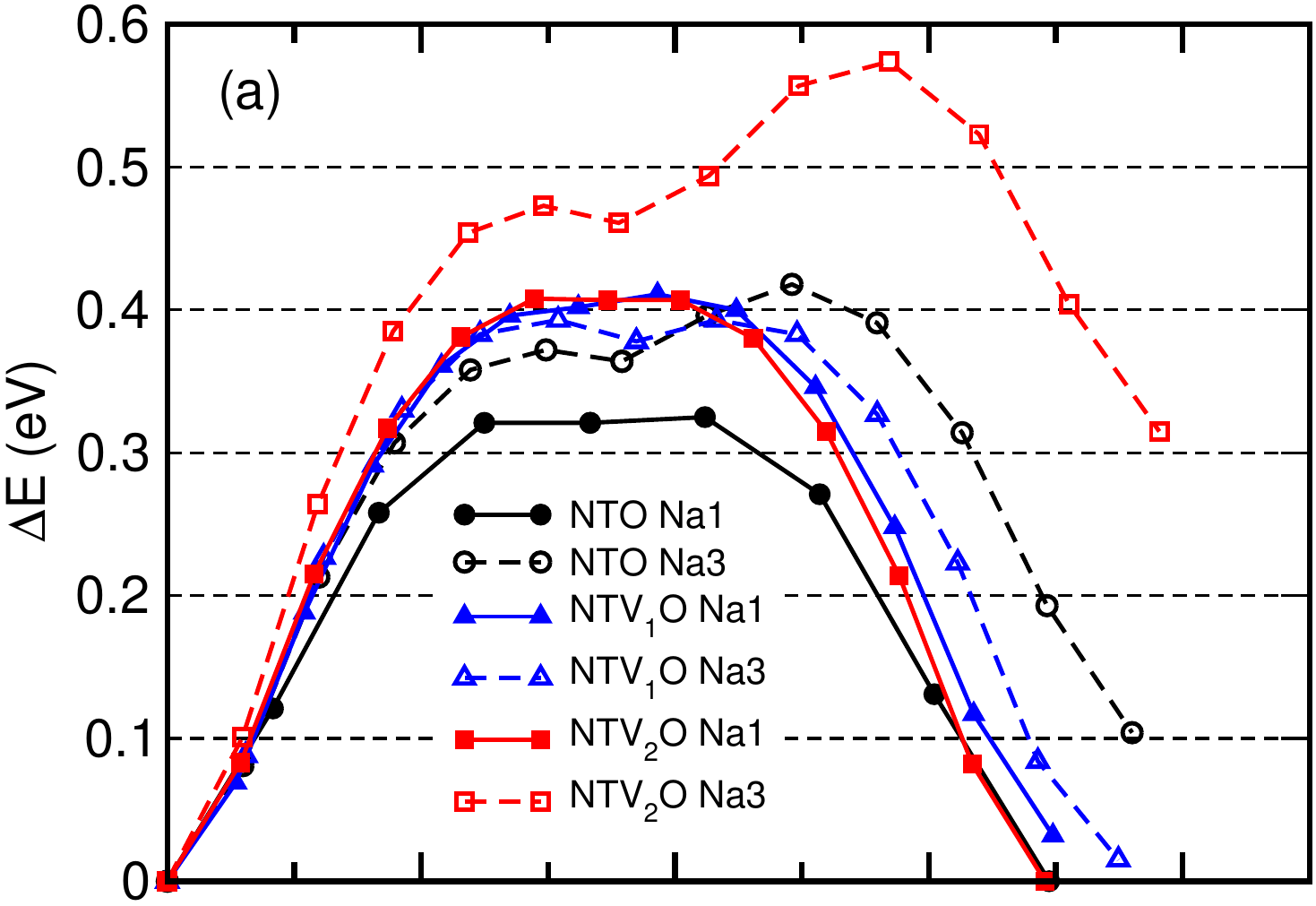} \\
\includegraphics[clip=true,scale=0.5]{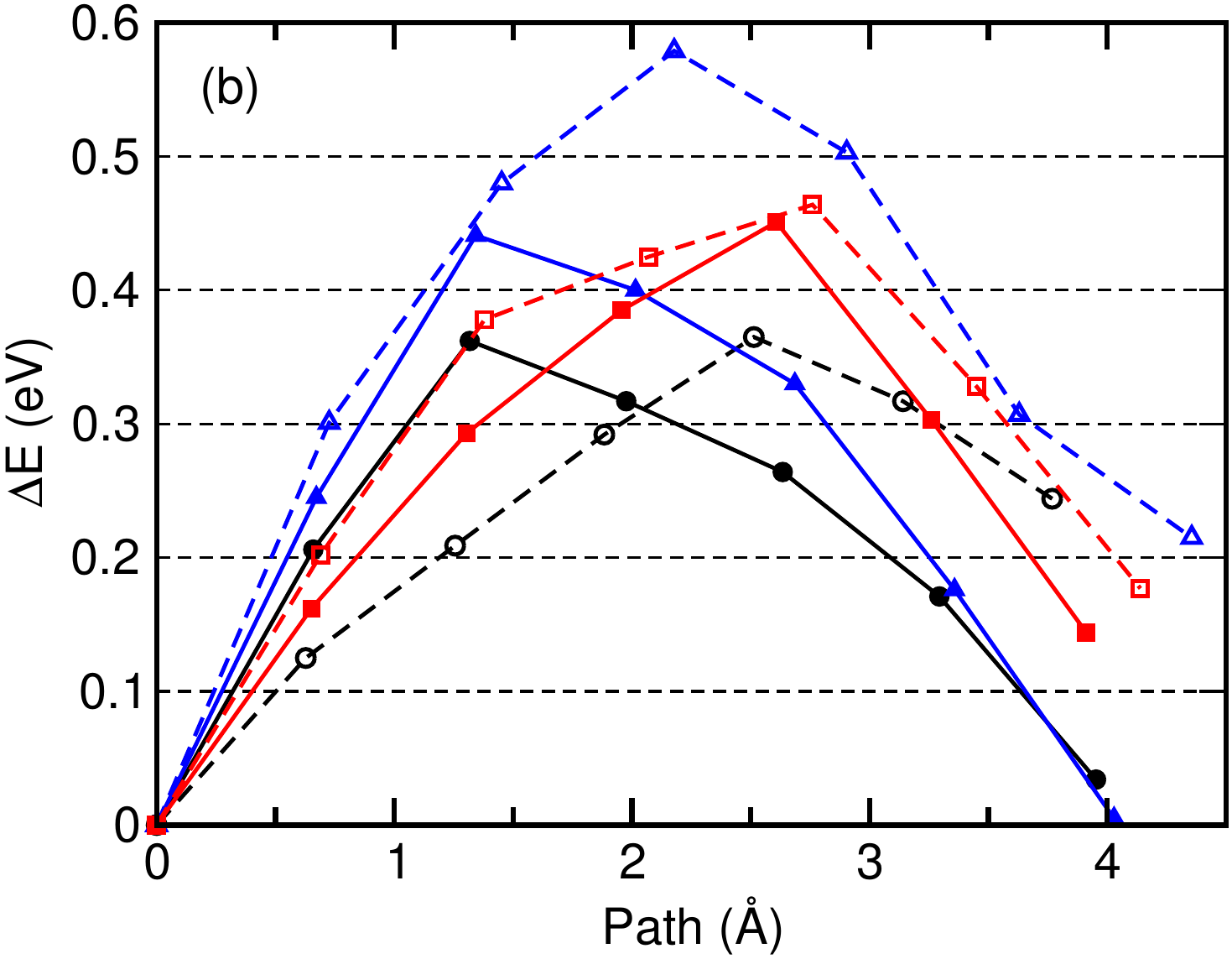}
\caption{Energy profile calculated with the NEB method for the migrations of Na ions along (a) (010) direction wavy-line pathway in sequential move way and (b) (100) direction step-line-step pathway in concurrent move way, depicted in Fig.~\ref{fig5}. Two different Na concentrations, one Na atom inserted into \ce{Na2M3O7} supercells (denoted as Na1 with solid line) and three inserted Na atoms (Na3 with dashed line) are considered for NTO, \ce{NTV1O} and \ce{NTV2O}.}
\label{fig6}
\end{figure}

To increase the reliability of ionic conductivity test, we considered two different Na concentrations, one Na atom inserted into \ce{Na2M3O7} supercells (\ce{Na_{2.25}M3O7}, denoted as Na1) and three inserted Na atoms (\ce{Na_{2.75}M3O7}; Na3).
Figure~\ref{fig5} presents the ball-and-stick view of Na ion migrations in the case of \ce{Na_{2.25}TiV2O7} (similar pathways were constructed for NTO and \ce{NTV1O} with two different Na concentrations).
Here, one cycle of Na ion migration consists of two processes; firstly the host Na ion at N$_{\text{h}}$ position moves toward vacancy at N$^1_{\text{i}}$ position (represented by green-colored balls) and then the inserted Na ion at N$^2_{\text{i}}$ position moves toward the generated vacancy at N$_{\text{h}}$ position (represented by orange-colored balls).
This is called sequential move way.
In the case of (100) direction, interestingly, although we let two Na ions move along the step-line-step pathway in the sequential way, they were observed in the NEB simulation to move simultaneously for all the cases of compounds.
This can be called concurrent move way, which was already found in the cathodic material \ce{NaFe(SO4)2}~\cite{yucj17pra}.

\begin{table}[!b]
\small
\caption{Activation energy values ($E_a$) for Na ion migrations along the (010) sequential and (100) concurrent pathways, and corresponding Na ion diffusion coefficient $D$ at room temperature.}
\label{tabl2}
\begin{tabular}{l@{\hspace{5pt}}c@{\hspace{5pt}}c@{\hspace{5pt}}c@{\hspace{5pt}}c@{\hspace{5pt}}c@{\hspace{5pt}}c}
\hline
 &    & \multicolumn{2}{c}{(010)} & & \multicolumn{2}{c}{(100)} \\
\cline{3-4} \cline{6-7}
Compound & & $E_a$ (eV) & $D$ (cm$^2$/s) & & $E_a$ (eV) & $D$ (cm$^2$/s) \\
\hline
NTO        & Na1 & 0.325 & 5.96$\times$10$^{-10}$ & & 0.362 & 1.44$\times$10$^{-10}$ \\
           & Na3 & 0.418 & 1.67$\times$10$^{-11}$& & 0.365 & 1.28$\times$10$^{-10}$\\
\ce{NTV1O} & Na1 & 0.411 & 2.18$\times$10$^{-11}$& & 0.441 & 6.88$\times$10$^{-12}$\\
           & Na3 & 0.393 & 4.36$\times$10$^{-11}$& & 0.579 & 3.41$\times$10$^{-14}$\\
\ce{NTV2O} & Na1 & 0.408 & 2.45$\times$10$^{-11}$& & 0.451 & 4.69$\times$10$^{-12}$\\
           & Na3 & 0.574 & 4.13$\times$10$^{-14}$& & 0.464 & 2.84$\times$10$^{-12}$\\
\hline
\end{tabular} \\
\end{table}
Figure~\ref{fig6} shows the energy profiles computed with the NEB method for Na ion migrations along these pathways.
Based on the obtained activation energy $E_a$, the Na ion diffusion coefficient can be estimated by $D=a^2\nu\exp(-E_a/k_{\text{B}}T)$, where $a\approx 4$ \AA, $\nu\approx10^{11}$ Hz and $k_{\text{B}}T=0.026$ eV~\cite{Pan13aem}.
Table~\ref{tabl2} summarize the calculated activation energies and Na ion diffusion coefficients.
For the case of NTO, they were determined to be 0.325 and 0.362 eV in (010) sequential and (100) concurrent ways, and slightly increased at higher Na concentration, being agreed well with the experimental value of 0.39 eV~\cite{Dynarowska17ssi}.
It turned out that the Ti/V exchange causes a slight increase of activation energy.
This might be due to the aforementioned enhancement of cation$-$anion interactions by Ti/V exchange in NTO.
Nevertheless, the calculated activation energies were below 0.579 eV in all the cases, which are not so high compared with other experimental data for NTO (0.67 eV)~\cite{Dynarowska17ssi}.
Therefore, such slight worsening of ionic conductivity would not be said to spoil the enhancement of electrode performance by Ti/V exchange.

\subsection{Electronic transport properties}
\begin{figure*}[!t]
\centering
\includegraphics[clip=true,scale=0.09]{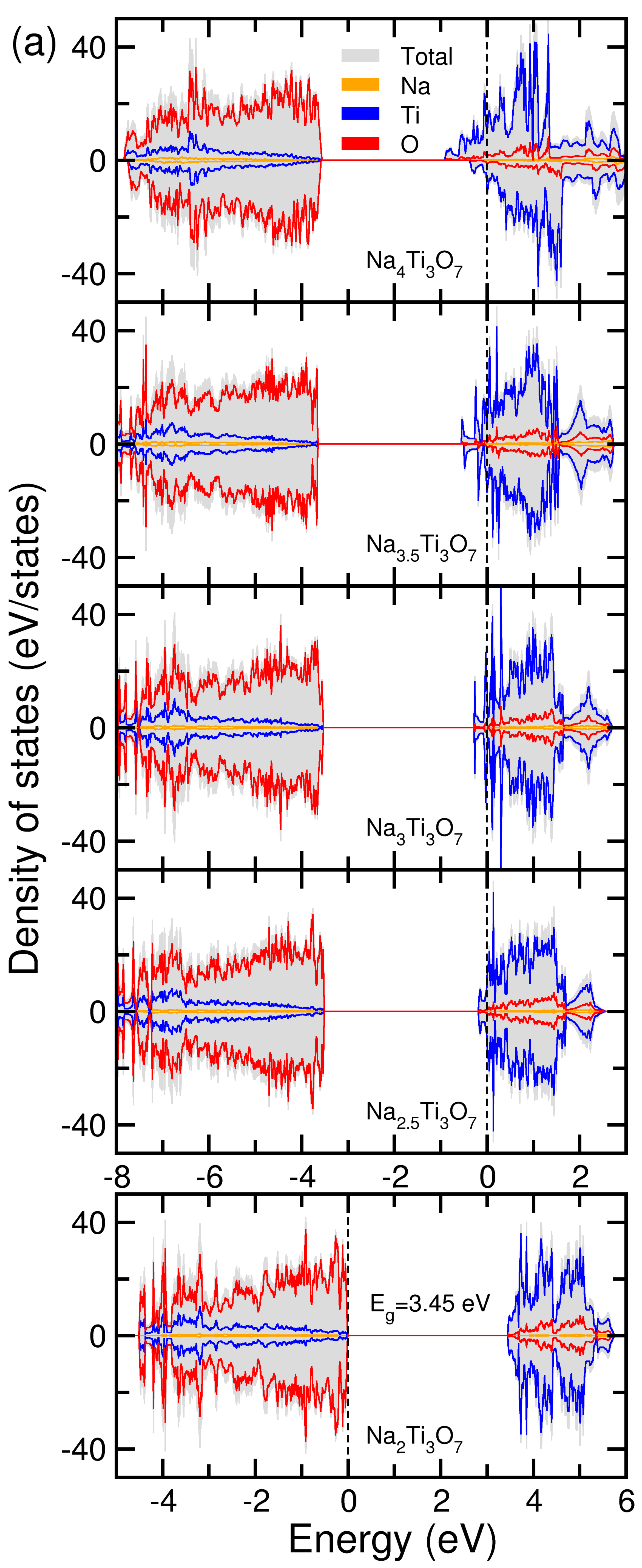}
\includegraphics[clip=true,scale=0.09]{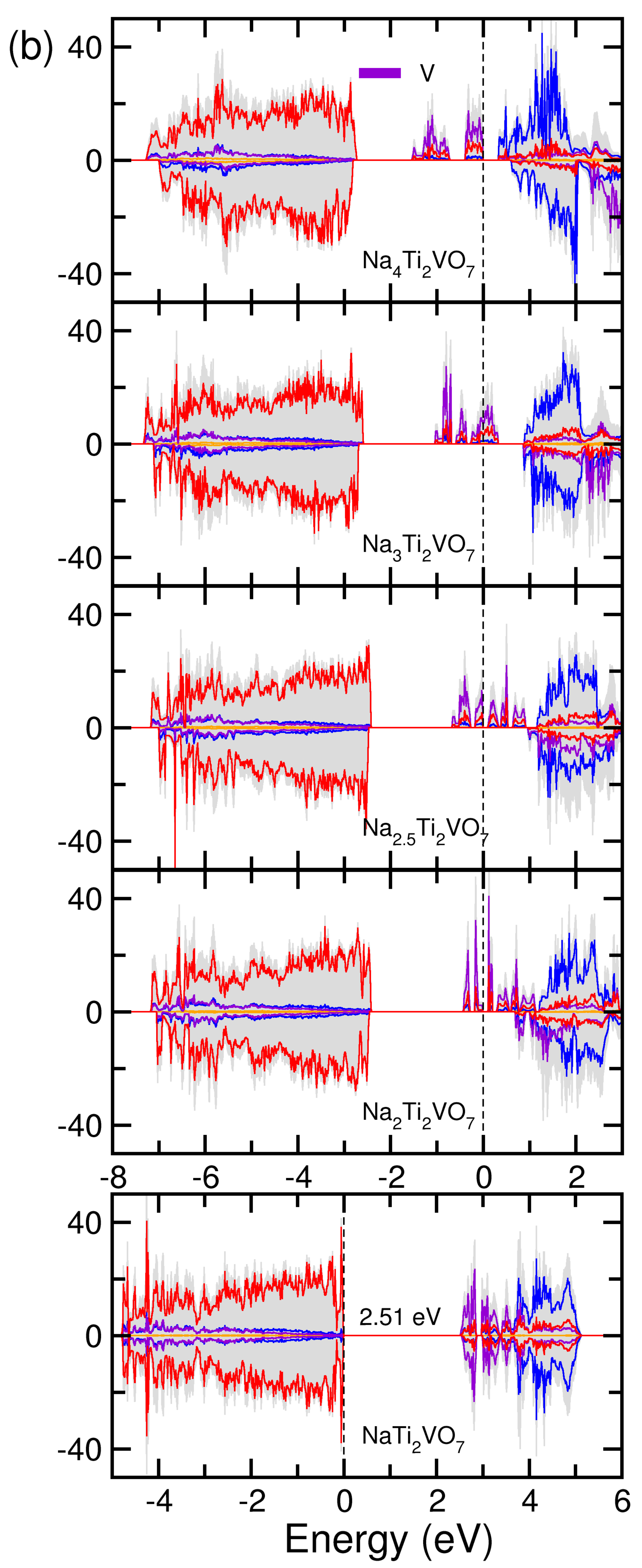}
\includegraphics[clip=true,scale=0.09]{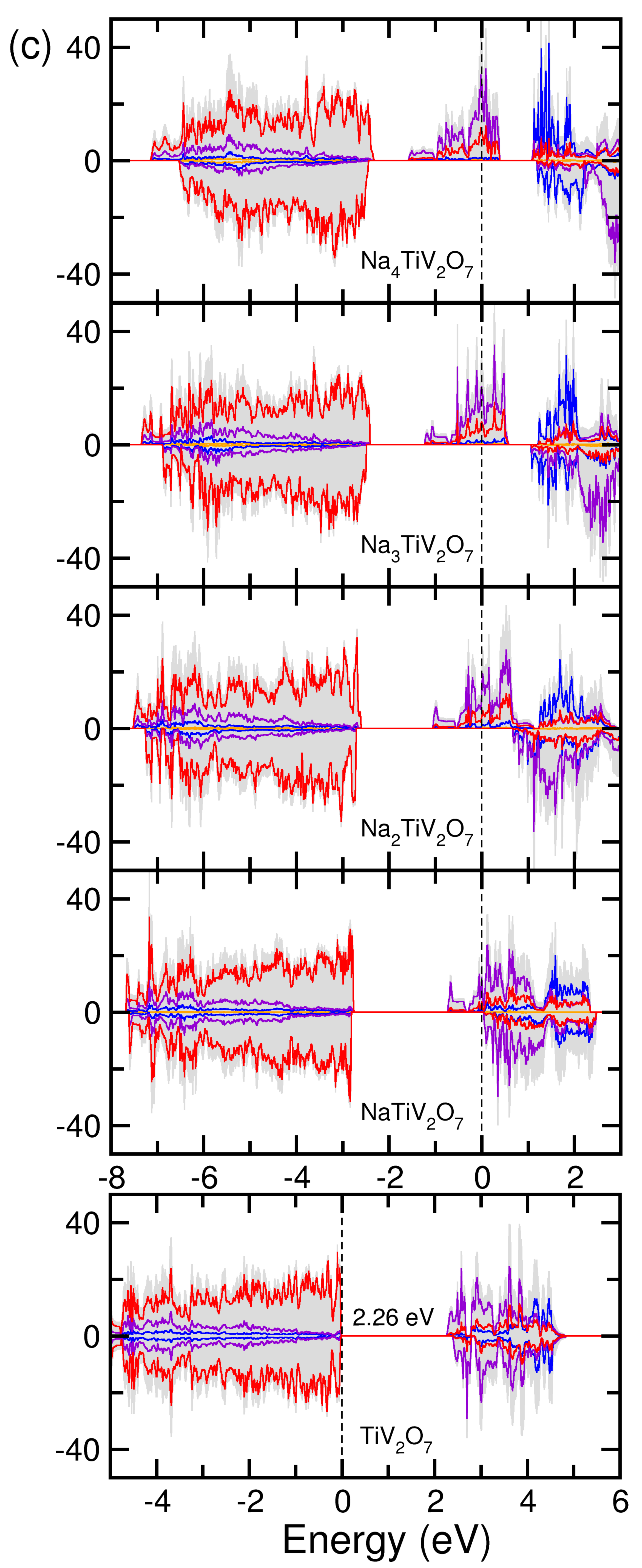}
\caption{Atomic resolved partial density of states in (a) \ce{Na_{2+$x$}Ti3O7} ($x=0, 0.5, 1, 1.5, 2$), (b) \ce{Na_{1+$x$}Ti2VO7} ($x=0, 1, 1.5, 2, 3$) and (c) \ce{Na_{$x$}TiV2O7} ($x=0, 1, 2, 3, 4$). Fermi level is set to zero and indicated by vertical dashed line.}
\label{fig7}
\end{figure*}

It was known that NTO is electrically insulating material with a direct bandgap of 3.73 eV~\cite{Pan13aem}, which is quite disadvantageous to electrode of batteries or capacitors.
What is the effect of Ti/V exchange on the electronic transport?
To answer this question, we investigated the atomic resolved partial density of states (PDOS) in the compounds under study, as shown in Fig.~\ref{fig7}.
The bandgap of \ce{Na2Ti3O7} was calculated to be 3.45 eV, which is very close to the experimental value of 3.73 eV~\cite{Pan13aem}, indicating that the GGA + $U$ method with $U=3$ eV for Ti and V is reasonable to give confidential results for electronic structure.
In this compound, the valence bands (VBs) close to the Fermi level ($E_{\ce{F}}$) were found to be dominated by O $2p$ states, while the conduction bands (CBs) to be characterized by Ti $3d$ states in partial hybridization with O $2p$ states.
To inquire into how the bandgap changes upon Ti/V exchange, we calculated the PDOS of \ce{NaTi2VO7} and \ce{TiV2O7}, in which Ti and V have the highest oxidation state of +4 and +5 (cf. Eqs~\ref{eq1b} and~\ref{eq1c}).
As indicated in Fig.~\ref{fig7}(b) and (c), the bandgaps were found to decrease as increasing the V content, such as 2.51 eV in \ce{NaTi2VO7} and 2.26 eV in \ce{TiV2O7}, revealing that Ti/V exchange can improve the electronic transport of electrode materials.
In these compounds, the VBs are governed by O $2p$ electrons as well, but the CBs are dominated by V $3d$ states with a minor contribution from O $2p$ states.

When inserting a Na atom into these base compounds, the Fermi level was found to move upward and be placed on the CBs consistently in the three kinds of models, resulting in change of electric state of material from insulating to metallic.
As already revealed in our previous work~\cite{yucj17pra}, this might be associated with the ionization of the inserted Na atom.
The electronic bands around the Fermi level were found to be formed mainly from Ti $3d$ spin-up and -down electrons for NTO and V $3d$ spin-up electrons for NTVO models with O $2p$ electrons.
Both the occupied and empty states of Na were observed far from the Fermi level, indicating that the inserted Na atom is fully ionized and the released electron moves to the Ti or V $3d$ and O $2p$ states around the conduction band minimum.
As a result, the Fermi level was pushed ahead, and the impurity-type bands were formed around it.
When increasing the amount of inserted Na atom, the occupied bands below $E_{\ce{F}}$ were found to gradually increase and the gap between the valence band maximum and conduction band minimum to get closer.

\begin{figure*}[!t]
\centering
\includegraphics[clip=true,scale=0.13]{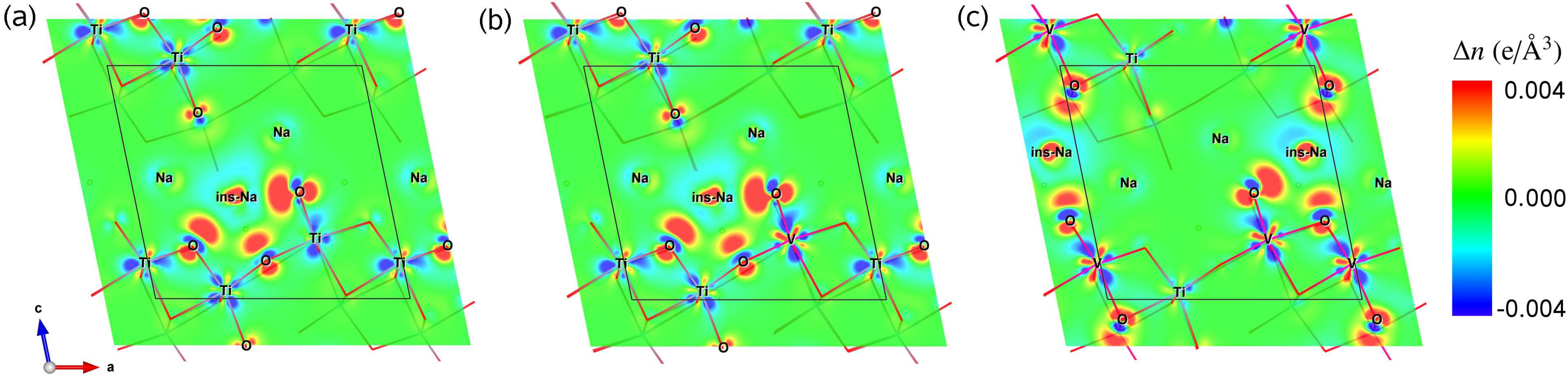}
\caption{Electronic density difference in (a) \ce{Na_{2.25}Ti3O7}, (b) \ce{Na_{2.25}Ti2VO7} and (c) \ce{Na_{2.25}TiV2O7}. Positive and negative values indicate electron accumulation and depletion respectively. The inserted Na atom is denoted as ins-Na.}
\label{fig8}
\end{figure*}
To clarify the electron transfer upon insertion of Na atom, we investigated the electronic charge density difference defined as,
\begin{equation}
\Delta n= n_{\ce{Na_{2.25}M3O7}}-(n_{\ce{Na2M3O7}}+0.25n_{\ce{Na}}), \label{eq4}
\end{equation}
where $n_{\ce{comp}}$ is the electron density of the compound.
Here, all the atomic positions were fixed in the same supercell dimension for the three compounds.
As shown in Fig.~\ref{fig8}, remarkable redistribution of electronic charge density was observed around the inserted Na atom and nearby Ti, V and O atoms.
One can see charge depletion around the Na atom, charge accumulation around the O atoms, and both accumulation and depletion around Ti and V atoms.
It can be thought that at the formation of \ce{Na2M3O7} the Na, Ti and V atoms offer their $3s$ and $3d$ electrons to the O atoms, becoming \ce{Na+}, \ce{Ti^{4+}}, \ce{V^{5+}} and \ce{O^{2-}} ions, and upon insertion of Na atom the released electrons transfer to metal ions, becoming \ce{Na+}, \ce{Ti^{3+}} and \ce{V^{3+}} ions.
This indicates that Ti and V atoms act as redox couples of \ce{Ti^{4+}}/\ce{Ti^{3+}} and \ce{V^{5+}}/\ce{V^{3+}} upon insertion/desertion of Na atom.
Such electron transfer was found to occur more remarkably when performing Ti/V exchange, indicating the stronger ionic bonding between the cations and anions in NTVO models.

\section{Conclusions}
Using first-principles calculations, we have investigated the electrochemical properties of \ce{Na2Ti3O7} and its two different Ti/V exchange compounds \ce{Na2Ti2VO7} and \ce{Na2TiV2O7}, which are thought to be promising anodic materials for Na-ion batteries and capacitors.
The lowest energy configurations for the two different exchange models have been determined with the structural optimizations combined with the crystalline symmetry consideration.
Through the bond valence sum analysis, we have identified the plausible positions for Na atoms inserting into \ce{Na2Ti3O7}, from which the final sodiation compounds have been fixed to be \ce{Na4M3O7}.
We have found that as increasing the Na content $x$ in \ce{Na_{$x$}M3O7}, the supercell volumes increase first and decrease after $x=3.25$ with the maximum relative volume expansion rates below 3\%, revealing the shrinkage of cell volume by Ti/V exchange because of their enhanced M$-$O interaction.
Based on the convex hull plot of formation energies of \ce{Na_{2+x}M3O7}, lower electrode voltages have been obtained in Ti/V exchange models with the maximal specific capacity of about 175 mAh g$^{-1}$, and by considering the higher oxidation state of \ce{V^{+5}}, the specific capacity could increase over 280 mAh g$^{-1}$.
We have computed the activation energies for Na ion migrations along the (010) direction wavy pathway in sequential move and (100) direction step-line-step pathway in concurrent move, finding that Ti/V exchange leads to slightly higher activation barriers.
Through the analysis of electronic density of states, we have found an improvement of electronic transport properties by Ti/V exchange in \ce{Na2Ti3O7}.

\section*{Acknowledgments}
This work is supported as part of the basic research project ``Design of Innovative Functional Materials for Energy and Environmental Application'' (No. 2016-20) by the State Commission of Science and Technology, DPR Korea.
Computation was done on the HP Blade System C7000 (HP BL460c) that is owned by Faculty of Materials Science, Kim Il Sung University.

\section*{\label{note}Notes}
The authors declare no competing financial interest.

\bibliographystyle{elsarticle-num-names}
\bibliography{Reference}

\end{document}